\documentclass[conference]{IEEEtran}
\IEEEoverridecommandlockouts
\usepackage{cite}
\usepackage{amsmath,amssymb,amsfonts}
\usepackage{algorithmic}
\usepackage{graphicx}
\usepackage{textcomp}
\usepackage{xcolor}
\usepackage{url}
\usepackage{pdflscape}
\usepackage{stfloats}
\usepackage{float}
\usepackage{placeins}
\usepackage{amsmath}
\usepackage[normalem]{ulem}
\def\BibTeX{{\rm B\kern-.05em{\sc i\kern-.025em b}\kern-.08em
    T\kern-.1667em\lower.7ex\hbox{E}\kern-.125emX}}
\begin{document}


\title{On the Contribution of Lexical Features to Speech Emotion Recognition}

\author{\IEEEauthorblockN{David COMBEI}
\IEEEauthorblockA{\textit{Computer Science Department} \\
\textit{Technical University of Cluj-Napoca, Romania} \\
david.combei@cs.utcluj.ro}
}

\IEEEoverridecommandlockouts
\IEEEpubid{\makebox[\columnwidth]{979-8-3315-3997-9/24/\$31.00~\copyright2025 IEEE \hfill} 
\hspace{\columnsep}\makebox[\columnwidth]{ }}

\maketitle


\begin{abstract}
Although paralinguistic cues are often considered the primary drivers of speech emotion recognition (SER), we investigate the role of lexical content extracted from speech and show that it can achieve competitive—and in some cases higher—performance compared to acoustic models. On the MELD dataset, our lexical-based approach obtains a weighted F1-score (WF1) of 51.5\%, compared to 49.3\% for an acoustic-only pipeline with a larger parameter count. Furthermore, we analyze different self-supervised (SSL) speech and text representations, conduct a layer-wise study of transformer-based encoders, and evaluate the effect of audio denoising.
\end{abstract}

\section{Introduction}
Speech Emotion Recognition (SER) has gained considerable interest in both research and industry, driven by its potential to improve human-computer interaction and health analytics \cite{b1}. SER applications range from emotion-aware virtual assistants and customer service bots to clinical monitoring and personalized user interfaces \cite{b2}. SER approaches rely on paralinguistic (acoustic) information, on how something is said, rather than the explicit lexical content of an audio sample. A variety of acoustic features are extracted from speech signals using classic speech processing techniques to serve as training or evaluation data for a machine learning task. These include acoustic features like fundamental frequency, energy, vocal quality indicators, formant frequencies and frequency-domain representations such as Mel-Frequency Cepstral Coefficients (MFCCs) or Linear Predictive Cepstral Coefficients (LPCC) \cite{b3,b4}.
While paralinguistic features are considered to be crucial, relying only on acoustics has its limitations. Studies have shown that acoustic-based SER models often yield suboptimal accuracy, especially in real-life settings \cite{b5, b6, b7}. In this work, we  challenge the long-standing paralinguistic paradigm which held that “how you say it” dominates “what you say” in detecting emotions using AI. In real-world conditions, both modalities carry weight. This paper embraces that perspective and aims to systematically examine the contributions of spoken words versus prosodic cues in emotion recognition. By leveraging natural emotional speech data, we seek to determine how much the words themselves contribute to detecting emotions, and whether relying only on lexical information can further push the boundaries of SER in real-world scenarios.
To summarize, our contributions are as follows:
\begin{itemize}
    \item We benchmark different SSL models.
    \item We show that the classifier does not need to be complex and the SSL derived representations are powerful enough.
    \item We systematically compare only lexical and only acoustic features in SER.
    \item We provide a quantitative analysis of how spoken words alone can contribute to emotion recognition performance in real-world conditions via automatic speech recognition (ASR).
    \item We benchmark a denoising neural network for both acoustic and lexical feature extraction.

\end{itemize}

\section{Related Work}
Recent works have leveraged advanced architectures capturing speech-based emotional cues. Chen et al. proposed SpeechFormer\cite{b9}, a transformer framework explicitly incorporating hierarchical attention in speech by processing frames, phonemes, words, and utterances in succession while keeping temporal information across them, which enabled fine-grained attention modeling and significantly reduced computational costs compared to standard transformers at that time. Building upon this idea, SpeechFormer++\cite{b10} followed and further optimized hierarchical attention, being 60\% faster for paralinguistic tasks on MELD, demonstrating superior performance and inference time.
Other approaches have investigated deformable and dynamic attention mechanisms to better handle temporal variability in speech. Deformable Speech Transformer (DST) introduced by Chen et al. \cite{b6} utilized learnable window positioning to adaptively locate emotionally relevant regions within speech signals, improving robustness on MELD. Similarly, DWFormer by Chen et al. \cite{b8} proposed dynamic windowing strategies with cross-window information interaction, allowing more precise temporal localization of emotional expressions, which is crucial for the multi-turn dialogues in MELD.

More recently, self-supervised representations have become central in most audio tasks, SER being no exception. Zhao et al. introduced TF-Mamba \cite{b7}, which integrated temporal-frequency state-space modeling to jointly capture temporal and spectral emotional features from self-supervised representations and has an ablation study, comparing different front-ends. Their method outperformed conventional SER models in terms of efficiency and latency on MELD, but DST still yields the state-of-the-art score. These methods highlight the trend of designing efficient and expressive architectures that balance fine-grained temporal dynamics with practical inference speed which are key challenges posed by real-world datasets like MELD.
Beyond purely acoustic modeling, Atmaja et al. \cite{b5} conducted a comprehensive survey on bimodal emotion recognition, emphasizing the value of fusing acoustic and lexical features for MELD and similar datasets. This is aligned with older but foundational studies such as Rozgic et al. \cite{b3}, who demonstrated that combining speech acoustics with text transcriptions improves emotion classification performance on these datasets.

\section{Methodology}

\subsection{Dataset description}
The Multimodal EmotionLines Dataset (MELD) \cite{b11} has become a pivotal benchmark for emotion recognition research, providing multi-speaker conversations annotated with emotional states across text, audio, and video modalities. MELD enables evaluation of models in real-world conditions. It contains seven emotions (anger, disgust, fear, joy, neutral, sadness and surprise) assigned by majority voting out of five workers assigned to annotate the utterances. The average duration of an utterance is 3.59 seconds. MELD has three partitions: train, development, and test. In our experiments (Table \ref{tab:acoustic_layerwise} and Table \ref{tab:lexical_layerwise}) the reported results are on the development set and the best performing models on this partition were used to compute the final scores on the test one, with the results presented in Table \ref{tab:sota}.

\subsection{Self-supervised models}

Our approach relies on the self-supervised representations extracted from large-scale pre-trained models. For the acoustic modality, we employ \texttt{wav2vec 2.0} \cite{b12} family variants: \texttt{wav2vec2-xls-r-2b}\footnote{\url{https://huggingface.co/facebook/wav2vec2-xls-r-2b}}, \texttt{wav2vec-BERT 2.0} \cite{b13}, and \texttt{wav2vec2-SER} \footnote{\url{https://huggingface.co/r-f/wav2vec-english-speech-emotion-recognition}}. All are used in a frozen setup to act as robust feature extractors without updating their weights. These self-supervised speech models were pre-trained on thousands of hours of raw audio and are known to capture fine-grained phonetic and prosodic cues relevant for speech emotion recognition \cite{b23}.

For the lexical modality, we integrate a set of strong transformer-based language models to extract semantic and contextual information from transcriptions: \texttt{BERT-base}, \texttt{BERT-large} \cite{b14}, \texttt{XLM-RoBERTa} \cite{b15}, and \texttt{DeBERTa} \cite{b16}. These models are also frozen during the feature extraction phase, ensuring consistent and transferable text embeddings while reducing computational cost.

\subsection{Extracting lexical representations from speech}

To begin the lexical-based experimental analysis, we first convert speech into text using an ASR system. Specifically, we employ the \texttt{Whisper-large-v3} \cite{b17} model for transcription, as it provides robust performance in diverse acoustic conditions and multilingual settings. The resulting transcripts are then fed into the text SSL models to extract the lexical representations which are then fed into the classifier.

\subsection{Classifier and training setup}

For the purpose of comparing acoustic and lexical characteristics and experimenting with various self-supervised models, because the SSL-derived representations have been proven to be strong enough on their own \cite{b19, b21}, we adopt a simple multi-layer perceptron (MLP) with three hidden layers as our classifier. For the speech representations, we apply average pooling over the time frames, while for the lexical representations, average pooling is performed over the token dimension. This results in a fixed-size, $1 x features$ vector serving as the input to the MLP.
 
 The MLP was trained for 500 epochs with a learning rate of $3e^{-5}$ with the Adam optimizer in each scenario, irrespective of the SSL or modality.

\subsection{Layer-wise analysis}

To better understand how emotional information is learned across different layers of self-supervised models, we conduct a layer-wise analysis by extracting representations from each transformer block of the speech and lexical encoders. For the speech modality, we analyze hidden states from individual layers of \texttt{wav2vec2-xls-r-2b}, \texttt{wav2vec-BERT 2.0}, and \texttt{wav2vec2-SER}. For the text modality, we do the same for \texttt{BERT-base}, \texttt{BERT-large}, \texttt{XLM-RoBERTa}, and \texttt{DeBERTa}. Each layer’s output is average-pooled over time frames  or tokens and then used for training, development and evaluation on the same MLP. This systematic evaluation allows us to identify which layers are more effective for emotion recognition. By comparing performance across layers, we aim to see if the intermediate representations learned by self-supervised models are better aligned with our task.

\subsection{Data denoising}

Due to the noise present in the MELD dataset and to improve the robustness of speech emotion recognition to real-world noise conditions, we apply data denoising as a preprocessing step. Specifically, we adopt the  speech enhancement by using DEMUCS \cite{b18} as a denoising neural network, which performs end-to-end enhancement directly in the time domain. This approach is capable of suppressing background noise and reverberation while preserving speech quality. Enhanced speech signals are then fed into the self-supervised models to extract cleaner representations and extend the experiments using them.

\section{Results}

\subsection{Acoustic-based speech emotion recognition}

\begin{table*}[t]
\caption{Layer-wise results for acoustic-based speech emotion recognition on the \textbf{development} partition of MELD dataset, including denoising (columns with D -).Values are weighted F1 scores (WF1 $\uparrow$). Higher is better. Underlined = best per column; boxed = global best.}
\label{tab:acoustic_layerwise}
\centering
\begin{tabular}{c||c|c|c||c|c|c}
\hline
\textbf{Layer} & \textbf{W2V2-XLS-R-2B} & \textbf{W2V-BERT 2.0} & \textbf{W2V2-SER} & \textbf{D -W2V2-XLS-R-2B} & \textbf{D -W2V-BERT 2.0} & \textbf{D -W2V2-SER} \\
\hline
0 & 38.89 & 30.93 & 38.62 & 37.34 & 31.91 & 39.28\\
1 & 38.43 & 38.45 & 40.29 & 37.79 & 38.96 & 40.13\\
2 & 40.78 & 41.58 & 41.81 & 38.93 & 40.91 & 39.57\\
3 & 41.33 & 44.51 & 42.04 & 40.82 & 42.86 & 41.51\\
4 & 42.77 & 44.88 & 43.34 & 41.13 & 44.72 & 41.26\\
5 & 41.51 & 47.74 & 44.24 & 41.04 & 46.28 & 42.31\\
6 & 42.76 & 47.89 & 44.01 & 41.36 & 46.88 & \textbf{\uline{43.73}}\\
7 & 43.42 & 46.84 & 43.58 & 42.39 & 45.95 & 43.37\\
8 & 44.28 & 46.37 & 43.64 & 43.75 & 46.12 & 42.72\\
9 & 45.79 & 44.82 & 45.56 & 43.07 & 45.72 & 42.48\\
10 & 45.60 & 47.28 & 45.13 & 44.32 & 45.06 & 43.19\\
11 & 45.67 & 47.14 & 45.38 & 43.83 & 46.27 & 42.39\\
12 & 46.39 & 47.32 & 44.57 & 46.46 & \textbf{\uline{48.81}} & 43.42\\
13 & 47.11 & 47.82 & 46.02 & 45.47 & 47.76 & 42.78\\
14 & 45.72 & \textbf{\uline{48.00}} & \textbf{\uline{46.59}} & 45.40 & 45.73 & 42.81\\
15 & 46.30 & 47.07 & 45.57 & 45.13 & 45.37 & 41.71\\
16 & 46.82 & 47.41 & 44.07 & 45.65 & 47.03 & 41.73\\
17 & 45.74 & 45.23 & 43.69 & 45.29 & 46.18 & 41.68\\
18 & 46.74 & 45.14 & 43.61 & 44.00 & 43.26 & 40.73\\
19 & 46.86 & 44.10 & 43.57 & 45.09 & 42.82 & 40.95\\
20 & 47.02 & 45.47 & 42.77 & 45.72 & 43.67 & 39.82\\
21 & 47.72 & 42.53 & 42.67 & 45.45 & 42.24 & 39.41\\
22 & 47.15 & 42.64 & 42.64 & 45.71 & 41.89 & 39.64\\
23 & 47.39 & 40.46 & 42.21 & 45.58 & 40.77 & 40.96\\
24 & 47.97 & 38.27 & 42.81 & 45.32 & 37.71 & 40.57\\
25 & 47.67 & - & - & 45.78 & - & -\\
26 & \boxed{\textbf{48.89}} & - & - & 45.55 & - & - \\
27 & 48.34 & - & - & 46.38 & - & - \\
28 & 47.52 & - & - & 45.63 & - & - \\
29 & 48.46 & - & - & 45.41 & - & - \\
30 & 47.83 & - & - & 46.15 & - & - \\
31 & 48.03 & - & - & \textbf{\uline{46.62}} & - & - \\
32 & 48.76 & - & - & 46.25 & - & - \\
33 & 48.42 & - & - & 46.25 & - & - \\
34 & 48.41 & - & - & 46.62 & - & - \\
35 & 47.83 & - & - & 46.3 & - & - \\
36 & 48.09 & - & - & 45.65 & - & - \\
37 & 47.55 & - & - & 45.49 & - & - \\
38 & 46.92 & - & - & 44.25 & - & - \\
39 & 46.59 & - & - & 44.88 & - & - \\
40 & 46.57 & - & - & 45.25 & - & - \\
41 & 44.92 & - & - & 44.3 & - & - \\
42 & 47.13 & - & - & 44.08 & - & - \\
43 & 45.12 & - & - & 45.08 & - & - \\ 
44 & 44.78 & - & - & 43.88 & - & - \\ 
45 & 44.60 & - & - & 44.65 & - & - \\
46 & 44.02 & - & - & 44.05 & - & - \\
47 & 43.85 & - & - & 43.31 & - & - \\
48 & 42.36 & - & - & 43.33 & - & - \\
\hline
\end{tabular}
\end{table*}

\begin{table*}[]
\setlength{\tabcolsep}{3pt}
\caption{Layer-wise results for lexical-based speech emotion recognition on the \textbf{development} partition of MELD dataset across various text SSL models, including denoising (columns with D -).Values are weighted F1 scores (WF1 $\uparrow$). Higher is better. Underlined = best per column; boxed = global best.}
\label{tab:lexical_layerwise}
\centering
\begin{tabular}{c|c|c|c|c|c|c|c|c}
\hline
\textbf{Layer} & \textbf{BERT-BASE} & \textbf{BERT-LARGE} & \textbf{XLM-ROBERTA} & \textbf{DEBERTA} & \textbf{D -BERT-BASE} & \textbf{D -BERT-LARGE} & \textbf{D -XLM-ROBERTA} & \textbf{D -DEBERTA} \\
\hline
0  & 43.54 & 44.78 & 44.03 & 43.93 & 43.02 & 41.84 & 40.75 & 41.55 \\
1  & 44.48 & 45.18 & 40.97 & 43.75 & 42.38 & 42.54 & 40.45 & 43.68 \\
2  & 43.87 & 44.63 & 42.15 & 45.13 & 41.99 & 41.40 & 42.15 & 42.50 \\
3  & 44.27 & 45.33 & 43.10 & 45.98 & 42.69 & 42.15 & 41.85 & 45.27 \\
4  & 45.29 & 46.07 & 42.84 & 47.09 & 43.37 & 42.59 & 42.49 & 44.52 \\
5  & 46.32 & 46.32 & 44.42 & 46.73 & 44.36 & 43.16 & 43.02 & 44.73 \\
6  & 45.90 & 46.34 & 43.28 & 47.47 & 43.43 & 43.79 & 43.53 & 45.59 \\
7  & 45.93 & 45.35 & 44.95 & 47.67 & 44.70 & 43.83 & 44.06 & 45.41 \\
8  & 47.40 & 45.54 & 44.55 & 47.37 & \textbf{\uline{46.63}} & 42.69 & 44.30 & 46.56 \\
9 & 48.57 & 45.67 & 45.10 & 47.72 & 46.46 & 44.38 & 43.39 & 46.37 \\
10 & \textbf{\uline{48.78}} & 46.34 & 44.73 & 47.31 & 46.44 & 44.69 & 43.37 & 46.01 \\
11 & 48.08 & 46.24 & 45.30 & 47.06 & 46.09 & 44.67 & 42.35 & 47.07 \\
12 & 47.98 & 47.23 & 45.75 & 48.26 & 45.66 & 44.22 & 43.96 & 46.47 \\
13 & -     & 46.21 & 45.56 & 49.54 & -     & 44.02 & 43.84 & 47.16 \\
14 & -     & 48.14 & 45.81 & 49.28 & -     & 45.21 & 45.07 & 47.35 \\
15 & -     & 48.01 & 46.15 & 50.56 & -     & 46.25 & 43.86 & 47.92 \\
16 & -     & 47.88 & 46.69 & 50.94 & -     & 46.28 & 45.27 & 48.79 \\
17 & -     & 49.29 & 47.16 & 50.40 & -     & \textbf{\uline{47.00}} & 45.05 & 48.65 \\
18 & -     & 49.50 & 46.35 & 49.91 & -     & 46.78 & 46.15 & 48.30 \\
19 & -     & \textbf{\uline{51.04}} & \textbf{\uline{47.48}} & \boxed{\textbf{51.73}} & -     & 46.73 & \textbf{\uline{47.09}} & \textbf{\uline{48.82}} \\
20 & -     & 49.57 & 46.60 & 51.01 & -     & 46.25 & 46.06 & 47.82 \\
21 & -     & 49.37 & 47.37 & 49.95 & -     & 46.76 & 45.85 & 48.68 \\
22 & -     & 49.80 & 46.48 & 49.40 & -     & 46.29 & 46.05 & 45.60 \\
23 & -     & 48.85 & 45.92 & 47.34 & -     & 45.63 & 44.37 & 44.83 \\
24 & -     & 48.40 & 44.17 & 47.40 & -     & 46.59 & 45.08 & 45.26 \\
\hline
\end{tabular}
\vspace{-.5cm}
\end{table*}

We evaluate the performance of the acoustic-based speech emotion recognition pipeline by using frozen self-supervised speech representations and training a fixed MLP classifier, as described in the previous section, on the training partition of the MELD dataset. Table \ref{tab:acoustic_layerwise} shows the layer-wise WF1 (column 1) across all transformer layers over the development partition of the MELD dataset. Additionally, we applied the same experimental pipeline to the denoised audio samples, performing layer-wise analysis for denoised \texttt{wav2vec2-xls-r-2b} (column 5), denoised \texttt{wav2vec-BERT 2.0} (column 6), and denoised \texttt{wav2vec2-SER} representations (column 7).

Interestingly, the mid-level representations of the \texttt{wav2vec2-xls-r-2b} model achieve the best performance, yielding a 48.9\% WF1 on top of the MLP. This result is on par in performance with the best-performing acoustic-based models from the literature, demonstrating that effective emotion recognition can be achieved even with simple classifiers when leveraging powerful self-supervised features.

Moreover, the observed peak performance at intermediate layers suggests that these layers capture more emotionally relevant phonetic features compared to final layers, where overly abstracted information is retained and fine-grained information might be lost. This trend aligns with prior findings in self-supervised speech learning, where early or mid-level transformer representations have been shown to encode information highly correlated with paralinguistic tasks \cite{b19, b21}.

Analyzing the impact of the denoising pre-processing step (columns marked with D- in Table~\ref{tab:acoustic_layerwise}) reveals a consistent degradation of performance across all acoustic models. Despite the MELD dataset containing noise, aggressive denoising using neural networks such as DEMUCS, appears to remove not only noise but also subtle non-verbal vocalizations that are crucial for emotion recognition. This outcome highlights that conventional speech enhancement techniques, while effective for other speech applications, might be detrimental for tasks like SER, where paralinguistic nuances matter. Therefore, task-specific noise reduction strategies or learning noise-robust representations directly within the model might be more beneficial.

\subsection{Lexical-based speech emotion recognition}

As a next step, we evaluate the potential of converging the speech emotion recognition task to a natural language processing (NLP) task using only lexical features by employing the same ideas as before, but instead of using the speech pre-trained models, the task will rely on text pre-trained auto-encoders. Although the dataset contains transcriptions, we would like to simulate real-world scenarios where no transcription is available, by passing the raw audio waveforms through \texttt{Whisper-large-v3} to obtain the transcriptions and then fed those into the auto-encoders to obtain the final representations followed by the classifier.

The results presented in Table \ref{tab:lexical_layerwise} demonstrates a clear trend: lexical-based models not only achieve competitive performance compared to acoustic-based systems but in many cases surpass them, with the best model reaching a WF1 of 51.73\% on the development set. This suggests that the semantic information captured by text representations can encode emotional relevant features with surprising results.

Another observation is that, similar to the acoustic models, intermediate layers of the text-based encoders tend to provide the highest performance. This indicates that emotional content is best captured when representations are not in the final layers where overfitting to general language modeling might arise, losing capability of capturing emotional relations.
Moreover, the lexical pipeline benefits from the robustness of modern ASR systems like Whisper, which enables high-quality transcriptions even under real-life scenarios, where no transcriptions are available. However, slight inaccuracies in transcription may introduce artifacts or errors in the entire pipeline.
To test this hypothesis, we perform emotion recognition using the transcripts in the metadata. The results show that, when using the manual transcriptions, the WF1 score is \textbf{60.9\%}. This result is highlighting a potential for future research on ASR error correction for speech emotion recognition tasks. 

Finally, the exploration of denoised audio feeding into ASR showed mostly degradation in most layers, likely due to distortions introduced by the denoising process. This reinforces the need for specialized enhancement models tailored for downstream tasks like SER, rather than generic speech enhancement (i.e. DEMUCS was pre-trained for music source separation).

\begin{table}[h]
\centering
\caption{Results on \textbf{test} partition of MELD for our top performing models for acoustic-based and lexical-based speech emotion recognition in comparison with state-of-the-art results. Best results are bolded}
\begin{tabular}{c|c|c}
\hline
\textbf{Model} & \textbf{Modality} & \textbf{WF1}\\
\hline
SpeechFormer\cite{b9} & Acoustic & 41.9 \\
Modality-Conversion\cite{b20} & Lexical & 43.1 \\
SpeechFormer++\cite{b10} & Acoustic & 47.0 \\
DWFormer\cite{b8} & Acoustic & 48.5 \\
TF-Mamba\cite{b7} & Acoustic & 48.5 \\
Deformable Speech Transformer (DST) \cite{b6} & Acoustic & 48.9\\
\hline
\textbf{Layer26 W2V2-XLS-R-2B + MLP (Ours)} & Acoustic & \textbf{49.3}\\
\textbf{Whisper + Layer19 DeBERTa + MLP (Ours)}  & Lexical & \textbf{51.5} \\
\hline
\end{tabular}
\label{tab:sota}
\end{table}

\subsection{Comparison with state-of-the-art}


Table \ref{tab:sota} summarizes our best-performing acoustic and lexical pipelines on the MELD test set against prior state-of-the-art systems. Our acoustic model (\textbf{Layer26 XLS-R-2B + MLP}) achieves \textbf{49.3\%} WF1, while our lexical system (\textbf{Whisper transcriptions + DeBERTa Layer19 + MLP}) reaches \textbf{51.5\%} WF1. 

These results highlight the potential to rethink SER pipelines by leveraging semantic content via ASR, enabling more efficient and effective emotion recognition that is deployable in real-world scenarios.

This approach has its own limitations, because on datasets such as RAVDESS\cite{b22}, where an utterance has multiple emotion audio files, the lexical-based approach won't work, since the classifier will learn a lot of noise due to multiple annotations on that utterance.

\section{Conclusions}

In this work, we systematically explored acoustic- and lexical-based speech emotion recognition on the MELD dataset, trying to identify if extracting lexical features from speech will provide any insights. Our results show that lexical content can carry substantial emotional information, with features extracted via ASR and text SSL, achieving superior performance (51.5\% WF1) compared to our best acoustic-based model (49.3\% WF1). While denoising and ASR are a great idea for this task, both need polishing to make them suitable. These findings suggest that treating speech emotion recognition as an NLP problem can unlock simpler and more powerful solutions in certain scenarios.

Future directions will include exploring ASR error correction and denoising strategies tailored for SER and a multimodal architecture with trainable modality-specific weights, allowing the model to learn how much to rely on acoustic versus lexical features. This will let us quantitatively assess the relative importance of each modality for emotion recognition.

Such efforts could lead to even more robust and practical emotion recognition systems suitable for real-world applications.

\section*{Acknowledgment}
This work was funded by the Romanian Ministry of Research, Innovation and Digitization project
DLT-AI SECSPP (id: PN-IV-P6-6.3-SOL-2024-2-0312).




\end{document}